\documentclass{aa}  

\usepackage{hyperref}

\usepackage{amsmath}
\usepackage{graphicx}
\usepackage{graphics}
\usepackage{subcaption}

\usepackage{float}

\usepackage{natbib}{}
\usepackage{xcolor}
\usepackage{url}

\bibpunct{(}{)}{;}{a}{}{,}

\def\teff{\ifmmode T_{\rm eff} \else $T_{\mathrm{eff}}$\fi}

\def\ltsima{$\buildrel<\over\sim$}
\def\lsim{\lower.5ex\hbox{\ltsima}}

\newcommand{\lya}{\ifmmode {\rm Ly}\alpha \else Ly$\alpha$\fi}

\newcommand{\ebv}{\ifmmode E_{\rm B-V} \else $E_{\rm B-V}$\fi}
\newcommand{\av}{\ifmmode A_{\rm V} \else $A_{\rm V}$\fi}

\def\msun{\ifmmode M_{\odot} \else M$_{\odot}$\fi}
\def\msunyr{\ifmmode M_{\odot} {\rm yr}^{-1} \else M$_{\odot}$ yr$^{-1}$\fi}
\def\zsun{\ifmmode Z_{\odot} \else Z$_{\odot}$\fi}

\def\lsun{\ifmmode L_{\odot} \else L$_{\odot}$\fi}

\def\mup{\ifmmode M_{\rm up} \else M$_{\rm up}$\fi}
\def\mlow{\ifmmode M_{\rm low} \else M$_{\rm low}$\fi}
\def\aap{A\&A}

\def\apjl{ApJL}
\def\apjs{ApJS}

\newcommand{\oh}{\ifmmode 12 + \log({\rm O/H}) \else$12 + \log({\rm
O/H})$\fi}
\newcommand{\ha}{\ifmmode {\rm H}\alpha \else H$\alpha$\fi}
\newcommand{\hb}{\ifmmode {\rm H}\beta \else H$\beta$\fi}
\newcommand{\hg}{\ifmmode {\rm H}\gamma \else H$\gamma$\fi}
\newcommand{\hd}{\ifmmode {\rm H}\delta \else H$\delta$\fi}
\newcommand{\he}{\ifmmode {\rm H}\epsilon \else H$\epsilon$\fi}
\newcommand{\oiii}{[O~{\sc iii}]}

\def\Oii{[O~{\sc ii}] $\lambda$3727}
\def\Oiii{[O~{\sc iii}] $\lambda\lambda$4959,5007}

\newcommand{\Neiii}{[Ne~{\sc iii}] $\lambda$3869}

\def\flyf{\ifmmode f_{\rm Lyf} \else $f_{\rm Lyf}$\fi}
\def\pz{\ifmmode P(z) \else $P(z)$\fi}
\def\ki2{\ifmmode \chi^2 \else $\chi^2$\fi}
\def\zphot{\ifmmode z_{\rm phot} \else $z_{\rm phot}$\fi}

\newcommand{\xphot}{\ifmmode x_\gamma \else $v_\gamma$\fi}
\newcommand{\xobs}{\ifmmode x_{\rm obs} \else $x_{\rm obs}$\fi}
\newcommand{\xcmf}{\ifmmode x_{\rm CMF} \else $x_{\rm CMF}$\fi}
\newcommand{\vexp}{\ifmmode V_{\rm exp} \else $V_{\rm exp}$\fi}
\newcommand{\vmax}{\ifmmode V_{\rm max} \else $V_{\rm max}$\fi}
\newcommand{\nh}{\ifmmode N_{\rm HI} \else $N_{\rm HI}$\fi}
\newcommand{\dv}{\ifmmode \Delta v({\rm em-abs}) \else $\Delta v({\rm em}-{\rm abs})$\fi}

\def\fesc{\ifmmode f_{\rm esc} \else $f_{\rm esc}$\fi}
\def\fescrel{\ifmmode f_{\rm esc,rel} \else $f_{\rm esc,rel}$\fi}

\def\frellya{\ifmmode f^{\rm rel}_{\rm{Ly}\alpha} \else $f^{\rm rel}_{\rm{Ly}\alpha}$\fi}

\newcommand{\mstar}{\ifmmode M_\star \else $M_\star$\fi}
\newcommand{\muv}{\ifmmode M_{1500} \else $M_{1500}$\fi}
\newcommand{\auv}{\ifmmode A_{\rm UV} \else $A_{\rm UV}$\fi}
\newcommand{\luv}{\ifmmode L_{\rm UV} \else $L_{\rm UV}$\fi}
\newcommand{\lir}{\ifmmode L_{\rm IR} \else $L_{\rm IR}$\fi}
\newcommand{\lbol}{\ifmmode L_{\rm bol} \else $L_{\rm bol}$\fi}
\newcommand{\liruv}{\ifmmode L_{\rm IR+UV} \else $L_{\rm IR+UV}$\fi}
\newcommand{\liroveruv}{\ifmmode L_{\rm IR}/L_{\rm UV} \else $L_{\rm IR}/L_{\rm UV}$\fi}
\newcommand{\nlyc}{\ifmmode N_{\rm Lyc} \else $N_{\rm Lyc} $\fi}
\newcommand{\rholyc}{\ifmmode \rho_{\rm Lyc} \else $\rho_{\rm Lyc} $\fi}
\newcommand{\chion}{\ifmmode \xi_{\rm ion} \else $\xi_{\rm ion}$\fi}
\newcommand{\chioncorr}{\ifmmode \xi_{\rm ion}^0 \else $\xi_{\rm ion}^0$\fi}

\begin{document}

\title{Strong Balmer break objects at $z \sim 7-10$ uncovered with JWST}
\subtitle{}
\author{A. Kuruvanthodi\inst{1},
D. Schaerer\inst{1,2}, 
R. Marques-Chaves\inst{1}, 
D. Korber\inst{1},
A. Weibel\inst{1},
P. A. Oesch\inst{1,3,4},
G. Roberts-Borsani\inst{1}
}
  \institute{Observatoire de Gen\`eve, Universit\'e de Gen\`eve, Chemin Pegasi 51, 1290 Versoix, Switzerland
\and CNRS, IRAP, 14 Avenue E. Belin, 31400 Toulouse, France
\and Cosmic Dawn Center (DAWN)
\and Niels Bohr Institute, University of Copenhagen, Jagtvej 128, DK-2200, Copenhagen N, Denmark
}

\authorrunning{Kuruvanthodi et al.}
\titlerunning{Strong Balmer break objects at $z \sim 7-10$ uncovered with JWST}

\date{Received date; accepted date}


\abstract{We report the discovery of robust spectroscopically confirmed Balmer break (BB) galaxies and candidates, with secure spectroscopic redshifts $7.1 \le z \le 9.6$ from publicly available \textit{James Webb} Space Telescope (JWST) extra-galactic photometric and spectroscopic surveys. To do so we used dedicated filters probing the Balmer break and inspected the objects with NIRSpec spectroscopy. We recover the previously known objects with strong Balmer breaks and reveal 10-11 new objects with clear BBs, thus tripling the number of spectroscopically confirmed galaxies with a BB at $z >7$. Approximately half of them show a pure BB and no signs of recent star formation, whereas the other half shows BB and emission lines, indicating most likely galaxies whose star formation ceased earlier and has restarted recently. 
Overall we find that $\sim 10-20$\% of all galaxies from our sample show signatures of an evolved stellar population. Furthermore, we find that the strength of the BB does not significantly depend on the rest-UV and rest-optical brightness of these sources.
In short, our work confirms that photometry alone has the potential to measure BB strengths and to identify evolved stellar populations at high redshift and that such objects may be more frequent than previously thought. The presence of galaxies with a range of break strengths and the joint presence of BB and emission lines indicate a bursty nature of the star formation in the early Universe.
}


\maketitle

\section{Introduction}
\label{s_intro}

The presence of Balmer break (BB) galaxies in the epoch of reionization has been proposed by multiple studies even in the pre-\textit{James Webb} Space Telescope (JWST) era \citep{Laporte2017Dust-in-the-Rei, Roberts-Borsani2020Interpreting-th, Hashimoto2018The-onset-of-st}. Historically, a strong color excess observed in Spitzer IRAC bands has been interpreted as BB for redshift $z \sim 6-7$ galaxies \citep[e.g.][]{eylesetal2007} which indicated a very high formation redshift for these sources. However, \citet{Schaerer_and_deBarros_2009_The_impact_of_nebular_emission} showed that this color excess could be due to strong emission lines like \oiii\ and thus not reflect a real BB. Multiple attempts were made to disentangle the true BB sources from emission line-dominated ones, by pushing to the high redshift so that these strong lines will be outside the IRAC bands \citep{Roberts-Borsani2020Interpreting-th, Hashimoto2018The-onset-of-st}. However, redshift uncertainties, the lack of photometric bands redward of the break, and the lack of spectroscopy made it hard to reveal the true nature of these sources.

This picture is changing the era of JWST with its remarkable sensitivity in the near-to-mid-infrared regime, availability of multiple medium bands, and spectroscopic capabilities. Today, already a few BB galaxies are spectroscopically observed above redshift $z>5$ \citep{Looser2024A-recently-quen, Strait_2023_An_Extremely_Compact, Witten_2024_Rising_from_the_ashes_A2744-YD4}, and more candidates have been proposed with photometry \citep{Laporte_2023_Resolving_ambiguities, Trussler_2024MNRAS_EPOCHS_IX_When_cosmic_dawn, Trussler_2024_Like_a_candle}. 

Detections of a Balmer break at high-$z$ indicate relatively advanced ages of stellar populations ($\ga 50$ Myr and older), probe the timescale of stellar mass assembly, and allow in principle to constrain the period of the onset of star-formation further in the past \citep[see e.g.][]{Mawatari2020Balmer-Break-Ga,Trussler_2024MNRAS_EPOCHS_IX_When_cosmic_dawn}. Furthermore, observations of BB galaxies and their statistics yield precious information on ceased or ``quenched'' star-formation in galaxies and constraints on the burstiness (intermittent star-formation) of galaxies and associated timescales 
\citep[cf.][]{Looser2024A-recently-quen,Witten_2024_Rising_from_the_ashes_A2744-YD4}.
It is therefore of great interest to identify new BB galaxies, determine their properties, and attempt to establish the statistics of these populations.

Here we report the spectroscopic discovery of $\sim 10$ new BB galaxies at $z \ga 7$, which almost triples the number of spectroscopically confirmed BB galaxies in the epoch of reionization. 

The observational data used for this study is described in Sect.~\ref{s_obs}. Our results are presented and discussed in Sect.~\ref{s_results}, and the main conclusions are summarized in Sect.~\ref{s_conclude}. Magnitudes are reported in the AB system, and we assume classical cosmological parameters ($H_{0} = 70$ km $s^{-1} Mpc^{-1}$, $\Omega_{M} = 0.3$, $\Omega_{\Lambda} = 0.7$).


\begin{figure*}[htb]
    \centering
    \includegraphics[width=18cm]{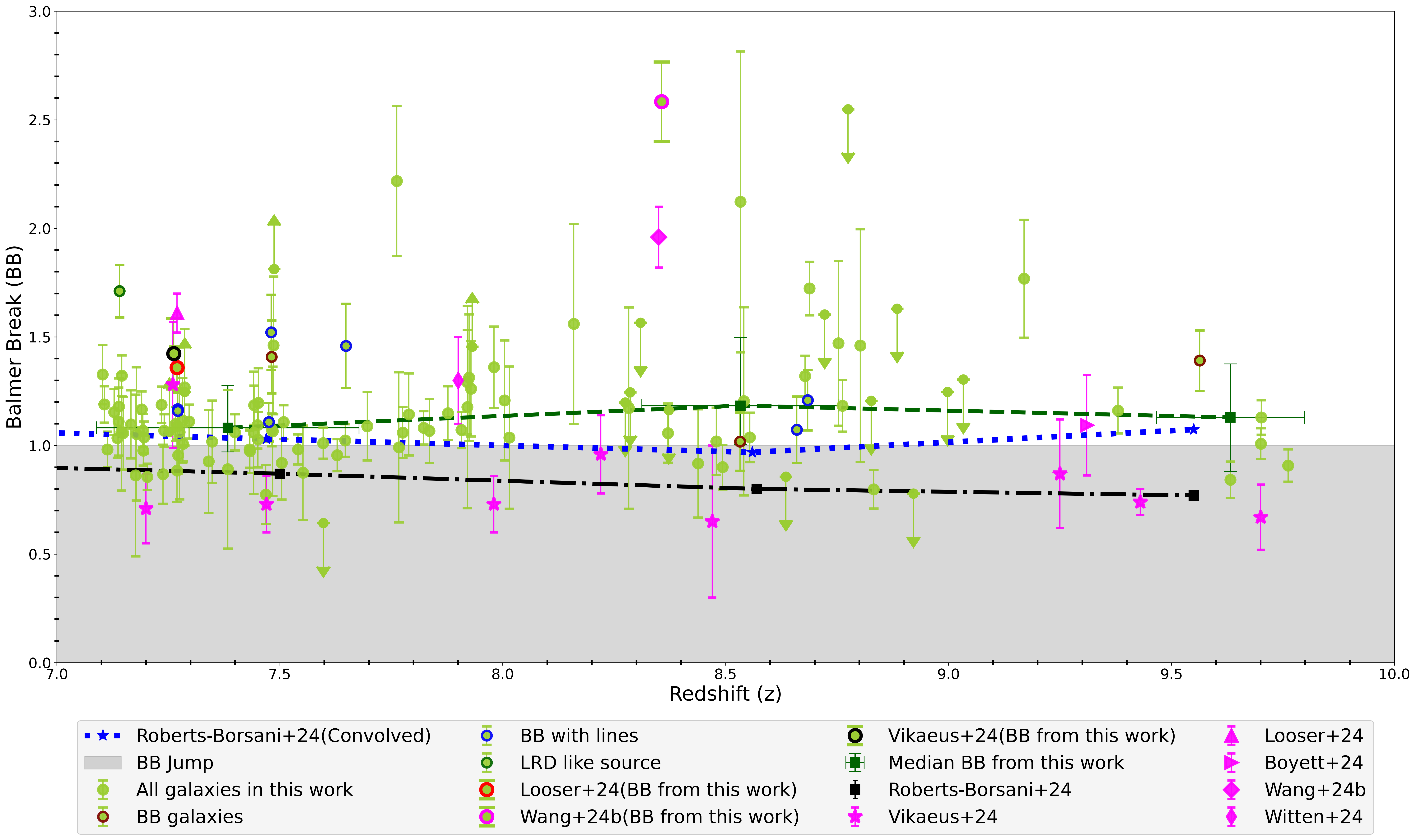}
    \caption{Balmer break strength (BB) measurements versus redshift for galaxies with secure spectroscopic redshifts. The grey area shows BB$\le 1$, often referred to as a Balmer jump. Green circles show the sources in this work and the green dashed line is the median BB obtained in three redshift intervals used in this work; all BB measurements are from photometry, and upper limits are $3 \sigma$. The BB measurements by \citet{Roberts-Borsani2024Between-the-Ext} using stacked JWST spectra are shown by the black dash-dotted line. For comparison, the blue dotted line shows our photometric BB measurements based on the same spectra, obtained from synthetic photometry using the same filters as for our sample. Balmer break objects from the literature are shown in magenta \citep{Vikaeus_2024MNRAS_To_be_or_not_to_be, Looser2024A-recently-quen, Boyett_2024NatAs_A_massive_interacting, Witten_2024_Rising_from_the_ashes_A2744-YD4}. 
    For the same objects we also show our BB  measurements from photometry (green circles with dark border).
    }
    \label{fig_bb_vs_z_lit}
\end{figure*}

\section{Observations and Balmer break measurements}
\label{s_obs}

\subsection{JWST NIRCam and NIRSpec observations}
To analyze the spectral energy distributions (SEDs) of high-$z$ galaxies and examine their Balmer breaks, we use publicly available multi-band imaging data from large extra-galactic surveys undertaken with the JWST \citep[CEERS, FRESCO, JADES, PRIMER;][Dunlop et al., in preparation]{Finkelstein_2023_CEERS_Key_Paper, Oesch_2023_The_JWST_FRESCO_survey, Eisenstein2023Overview-of-the}. In all these fields, NIRCam observations are available in the following seven filters, which we primarily use: F115W, F150W, F200W, F277W, F356W, F410M, F444W. Where available, we also use HST filters below 1$\mu$m (i.e. F435W, F606W, and F814W). We use the photometric catalogs constructed by \cite{Weibel_2024_Galaxy_build_up_in_the_first}, to which we refer for details regarding data reduction, source extraction, and flux measurements. Basically, we selected objects in the stacked F277W+F356W+F444W detection image. The 5-$\sigma$ depth in the F277W (F444W) band of the surveys considered here are 28.51 (28.21), 28.62 (28.36), 29.16 (28.84), and 29.93 (29.20) for PRIMER-UDS, PRIMER-COSMOS, CEERS (EGS), and FRESCO + JADES (GOODS-NORTH/SOUTH) respectively \citep[see ][]{Weibel_2024_Galaxy_build_up_in_the_first}. 
Furthermore, we require $\ge 3 \sigma$ detections in the filters required for the measurement of the BB (see Sect.~\ref{subsec_bb_from_phot}), which allows us to accurately determine BBs for galaxies down to $\sim 29$ magnitude in F444W.

To identify galaxies with secure spectroscopic redshifts we then used the DAWN JWST Archive \citep[DJA and msaexp][]{Heintz2024The-JWST-PRIMAL, brammer_2023_8319596}\footnote{\url{https://dawn-cph.github.io/dja}} retaining only galaxies with robust spectroscopic redshifts based on visual inspection of individual spectra (grade=3).
In the redshift interval $7.1 < z < 10$  we find 118 galaxies with matches in our photometric catalog, primarily taken from the above-cited surveys.

Since the spectra data originate from a variety of different programs, each with a priori different selection criteria and essentially unknown selection functions, it is not clear if our selection is representative of the average galaxy population at high-$z$, and rigorous statistical inferences cannot be derived from our work. To the best of our knowledge, none of the JWST programs used here targeted specifically objects with strong Balmer breaks. In this case, and if the selection criteria used by different programs are diverse enough, it is possible that the analyzed sample resembles somewhat a random selection. In any case, our analysis can at least be used to reveal new, previously unknown galaxies with Balmer breaks and to estimate the frequency of those objects among the available data.

\subsection{Balmer break measurements from photometry}
\label{subsec_bb_from_phot}
We then selected sources in different redshift bins where we could use NIRCam filters to probe the Balmer break. In practice, at redshifts $z=7.1-8.0$, $8.0-9.2$, and $9.2-10$ we use the following colors to measure the Balmer break: F277W-F356W, F277W-F410M, and F356W-F444W. With these redshifts and filter choices, the redward filters probe rest-wavelengths $>3500$ \AA\ and avoid strong emission lines of $\hb$ and \oiii. The above colors are then a good measure of the SED close to and across the Balmer break, as comparisons with spectroscopic measurements and simulations presented below show. The BB is measured for sources showing $\ge 3 \sigma$ detections in one or two of the BB-probing filters used, and upper or lower limits are reported for objects only detected in one of the filters. The rest of the sources with $< 3 \sigma$ in both BB-probing filters are ignored. For 101 objects (74, 21, and 6 in the three redshift bins) we obtain BB measurements; and for 17, upper or lower limits on BB.

The strength of the Balmer break (hereafter loosely referred to as ``Balmer break'' or BB ) of our sample is defined here as $BB = F_\nu^+/F_\nu^-$, where the flux densites longward and shortward are taken from photometry, and shown as a function of redshift in Fig.~\ref{fig_bb_vs_z_lit}. To compare our method with direct measurements from spectroscopy, we have convolved the stacked spectra from \citet{Roberts-Borsani2024Between-the-Ext} with the NIRCam filters used for our work to derive the corresponding photometric BB measurements. The results are also shown in Fig.~\ref{fig_bb_vs_z_lit} together with the BB measurements from the same spectra obtained by these authors. As expected, the photometric BB values are slightly higher than the spectroscopic ones (by $\sim 0.1-0.2$), since the filter redward of the break contains several weak emission lines (\Oii, \Neiii, and high-order Balmer lines). This broadly validates our method, showing that objects with $BB \ga 1.1-1.2$ should have true Balmer breaks. 

\begin{figure*}[htb]
    \centering
    \includegraphics[width=8.7cm]{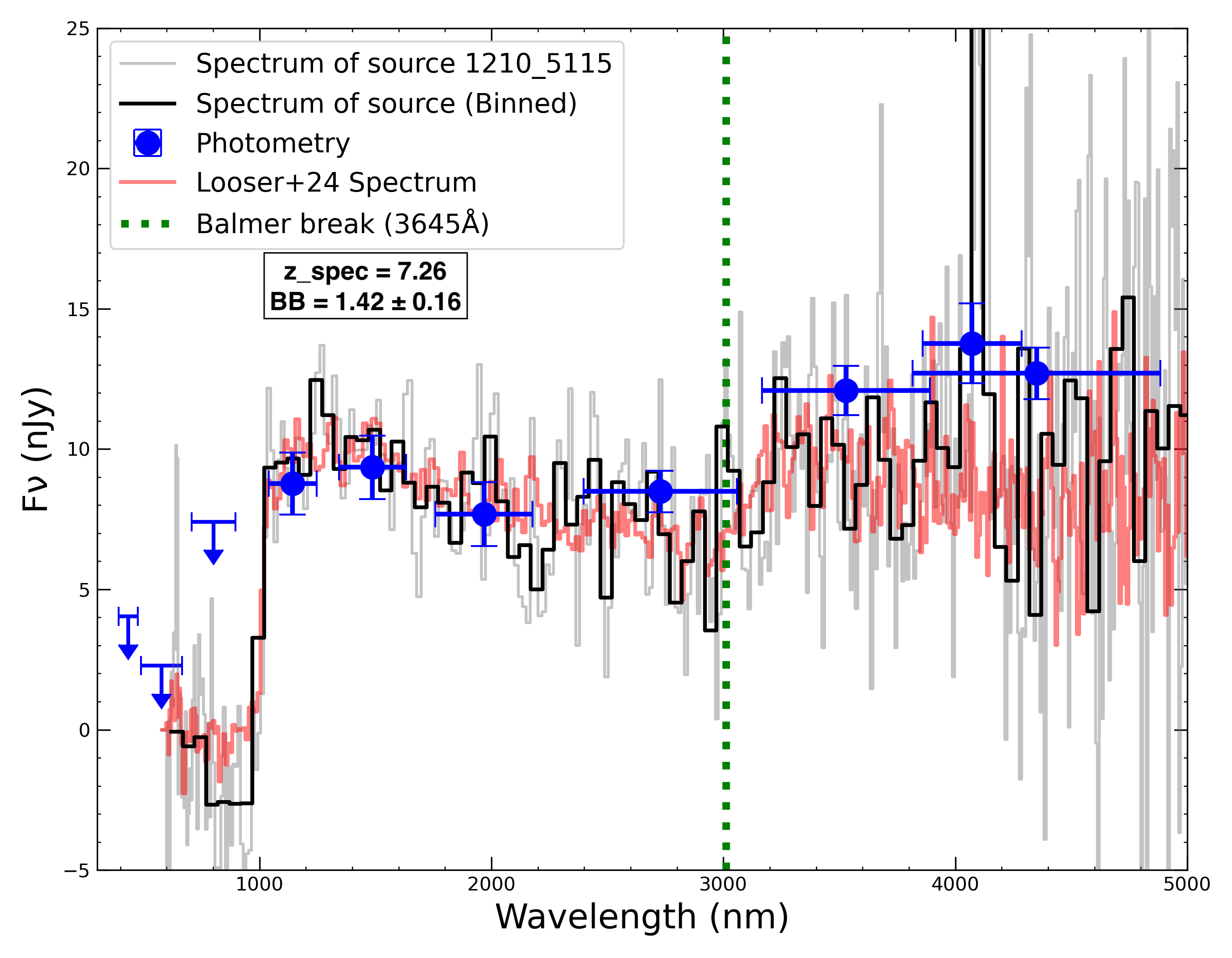}
    \includegraphics[width=8.7cm]{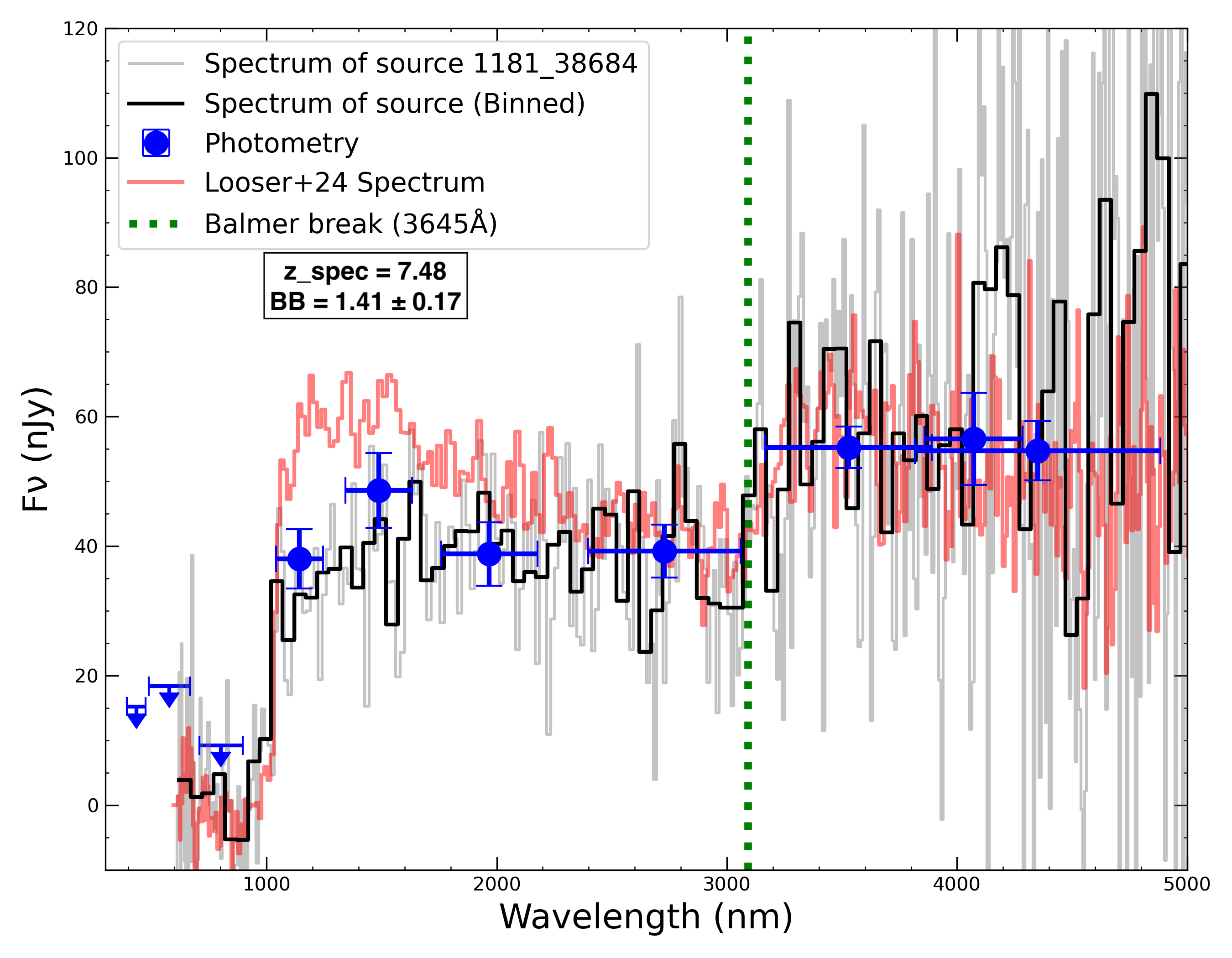}
    \includegraphics[width=8.7cm]{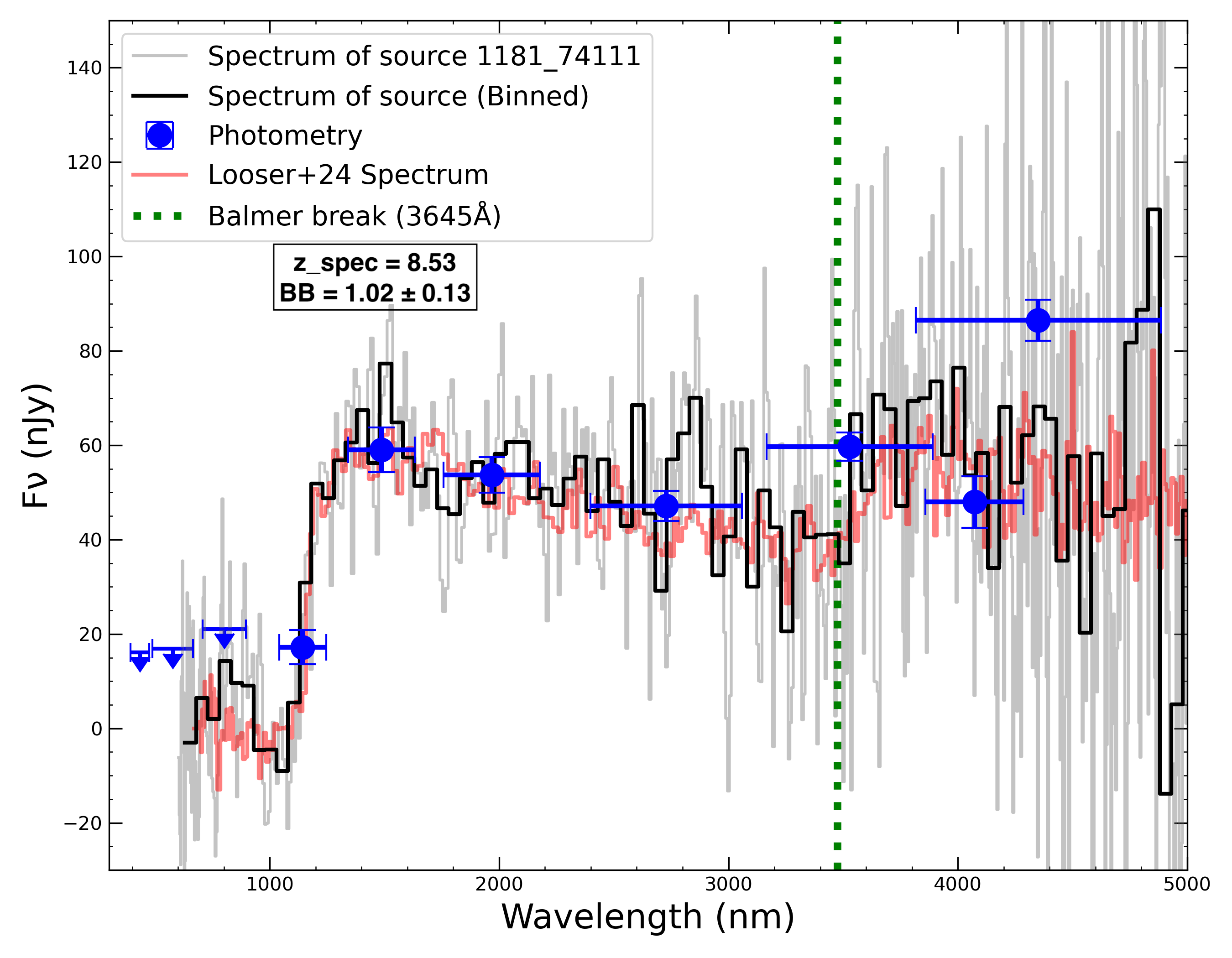}
    \includegraphics[width=8.7cm]{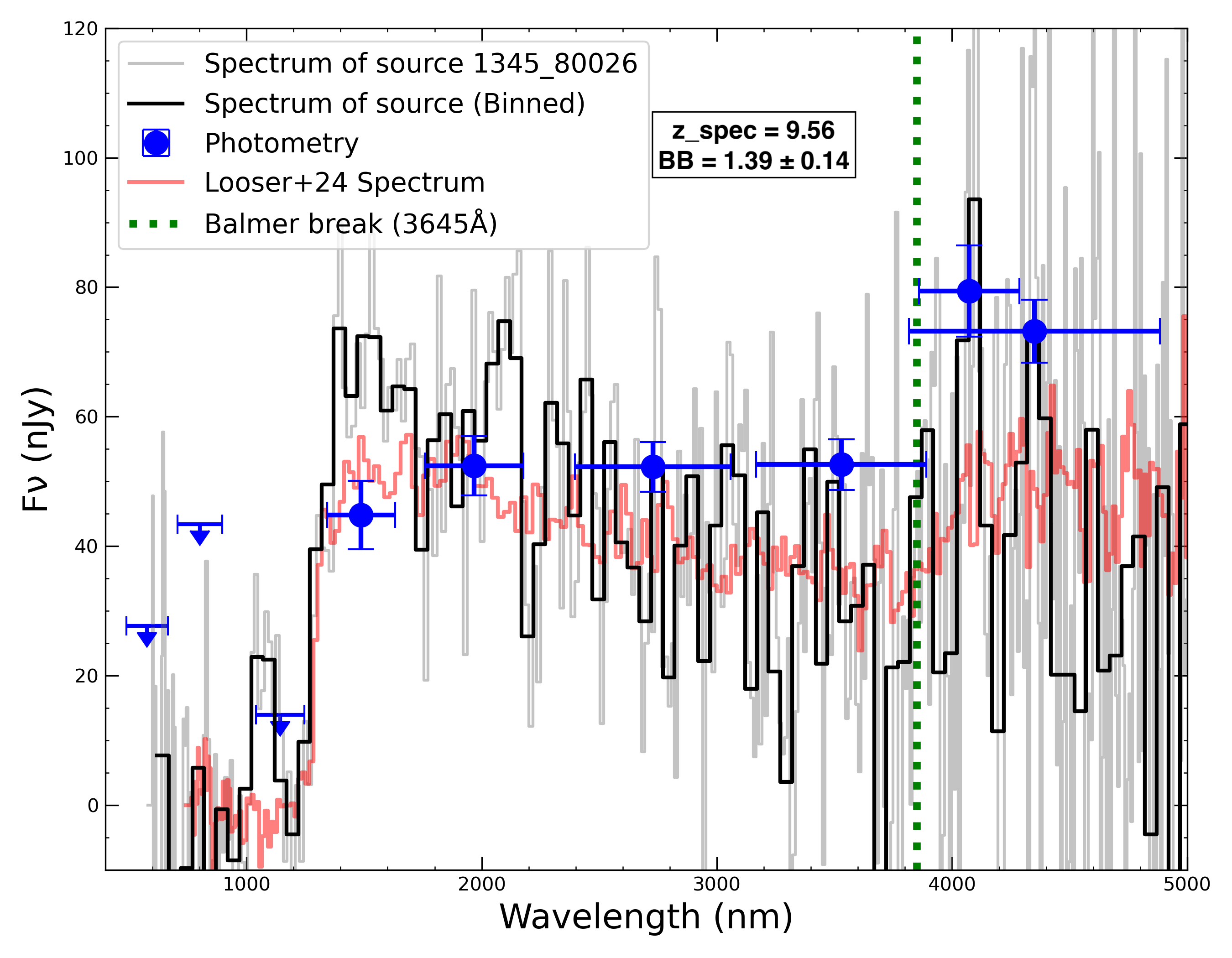}
    \caption{NIRSpec PRISM spectra of selected strong BB candidates at $z \sim 7-9.6$, showing objects with clear or plausible Balmer breaks. Binned and unbinned spectra are shown for illustration. For comparison we also overplot the spectrum of the mini-quenched $z=7.3$ galaxy from \cite{Looser2024A-recently-quen}, redshifted and rescaled to approximately match the observed flux blueward of the BB. \textbf{Top left:} Clear BB galaxy at $z=7.26$ (object in \cite{Vikaeus_2024MNRAS_To_be_or_not_to_be}. \textbf{Top right:} BB galaxy at $z=7.48$.
    \textbf{Bottom left:} BB candidate at $z=8.53$. \textbf{Bottom right:} BB candidate at $z=9.56$.}
    \label{fig_spectrum_CLEAR_BB_objects}
\end{figure*}

\section{Balmer breaks revealed in $z \sim 7-9$ galaxies}
\label{s_results}
The main result from our work, BB measurements as a function of redshift, is illustrated in Fig.~\ref{fig_bb_vs_z_lit}, together with BB measurements reported in the literature. First, we find a large number of individual galaxies at $z>7$, which show a significant BB. For example, 48 (27) galaxies out of 101 have photometric BB $>1.1$ (1.2). The strongest BB we measure is BB$\sim 2-2.5$ in two galaxies at $z=7.76$ and $z=8.35$, which surpass those reported earlier in the literature at $z \sim 7-7.5$, including e.g. the (mini)-quenched galaxy from \cite[][hereafter L24]{Looser2024A-recently-quen}. The BB of these objects are comparable to those reported by \citet[][hereafter W24]{Wang_2024_RUBIES_Evolved_Stellar_Populations} in three $z=6.9-8.4$ galaxies, including one object at $z=8.36$ for which they measure BB$=1.96 \pm 0.14$ from the NIRSpec spectrum.  
Second, at $z \ga 8.5$ and possibly up to $z=9.56$ we find several galaxies with fairly secure Balmer breaks, some confirmed by NIRSpec spectroscopy (see below). Finally, the median BB strength measured for our sample is in agreement with the measurements from the stacked JWST spectra of \cite{Roberts-Borsani2024Between-the-Ext}, showing that on average the Balmer break is close to unity from photometry, which is compatible with the presence of a small Balmer jump plus weak emission lines longward of the Break, as shown by these authors. However, photometry allows us to reveal the presence of a BB for individual objects for larger samples, and for objects beyond the reach of spectroscopy.

\begin{figure*}[htb]
    \centering
    \includegraphics[width=8.7cm]{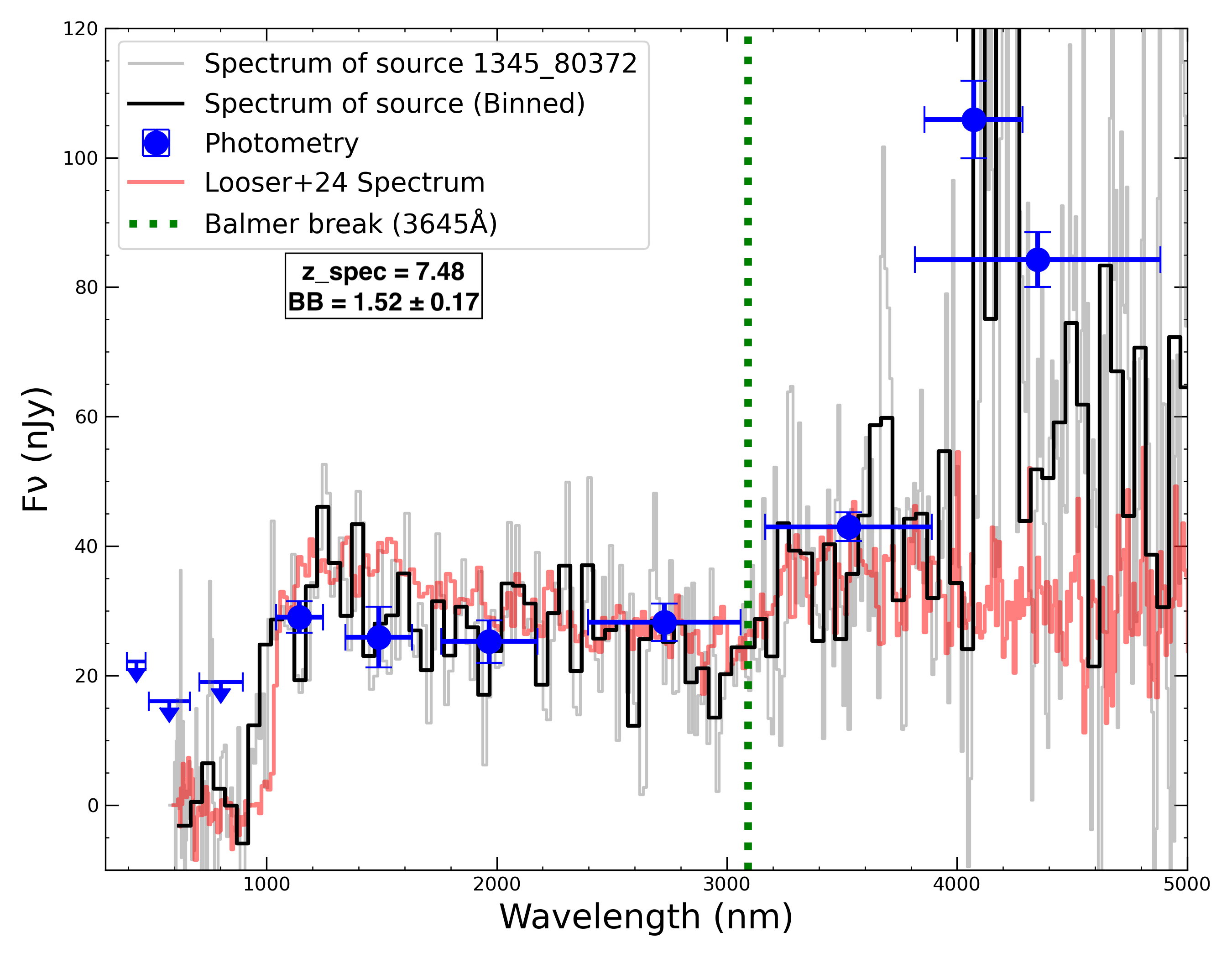}
    \includegraphics[width=8.7cm]{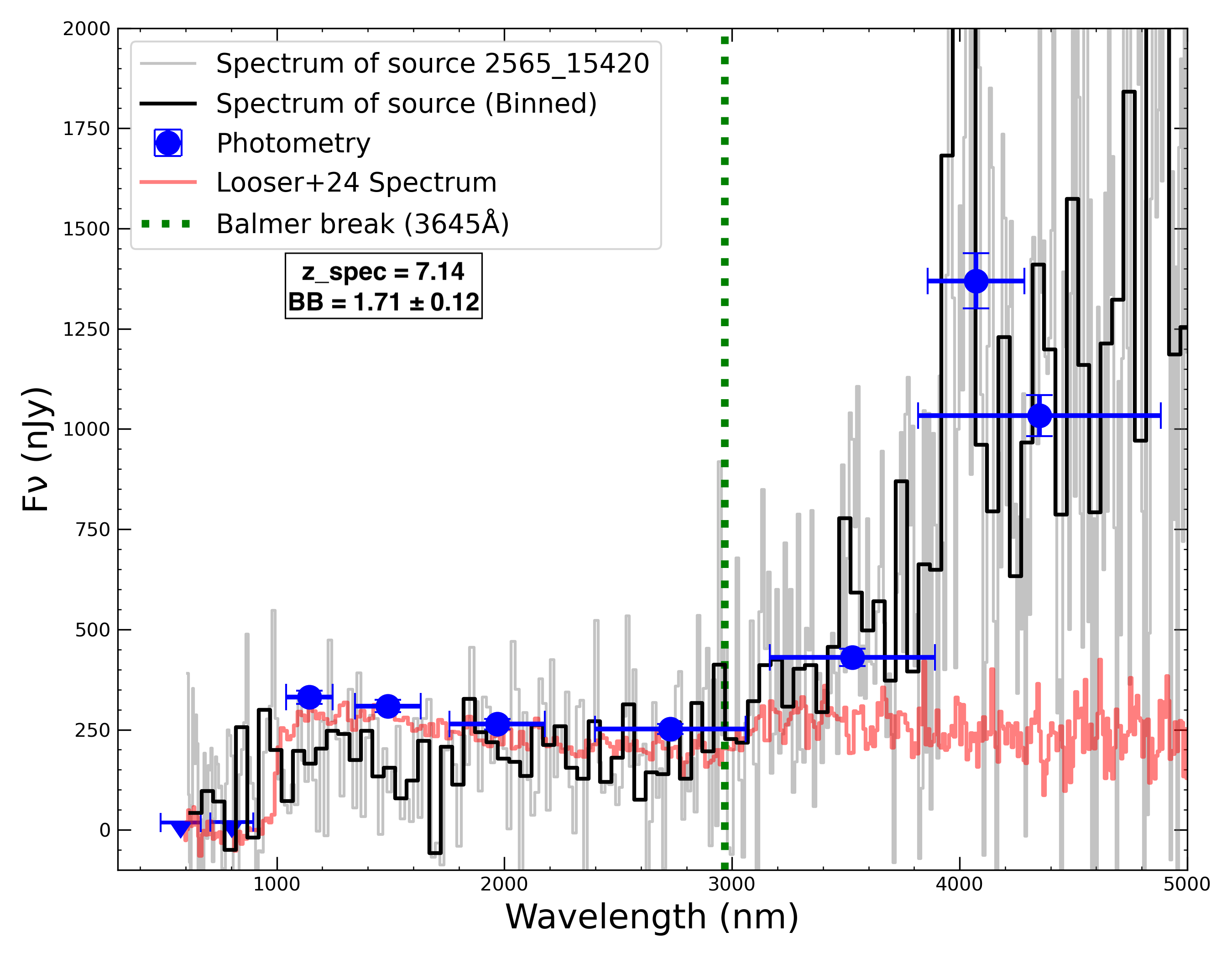}
    \caption{NIRSpec PRISM spectra of selected BB candidates with plausible Balmer breaks. Labels are the same as in Fig. \ref{fig_spectrum_CLEAR_BB_objects} \textbf{Left:} Clear BB galaxy at $z=7.48$ with emission lines, \textbf{Right:} Spectrum of an LRD candidate at $z=7.14$}
   \label{fig_spectrum_objects}
\end{figure*}


\subsection{Examination of the strong BB candidates in our sample}
Three of our 48 galaxies with BB$>1.1$ have previously been reported as galaxies at $z>7$ with Balmer breaks using spectroscopy. This includes the mini-quenched galaxy at $z=7.29$ from L24, object JADES 8079 from \citep[][hereafter V24]{Vikaeus_2024MNRAS_To_be_or_not_to_be}\footnote{These authors also report 8115 which is identical to the galaxy from L24.}, and an LRD candidate at $z=8.35$ from W24. The photometric and spectroscopic BB measurements of these objects are plotted in Fig. \ref{fig_bb_vs_z_lit}, showing good agreement for two of them, and a larger photometric BB for the W24 object, which is explained by the presence of lines and the relatively red spectrum longward of the break.

Although our BB measurements rely purely on photometry (and spectroscopic redshifts), we can use the available spectra to examine if our BB candidates are plausible or not. The NIRSpec PRISM spectra of selected strong BB candidates with BB$ \ga 1.4$ and their photometric fluxes are shown in Fig.~\ref{fig_spectrum_CLEAR_BB_objects} and \ref{fig_spectrum_objects}, and compared to the $z=7.3$ mini-quenched galaxy of L24. The spectra represent the following three categories found from inspection of the data: {\em 1)} Robust BB galaxies or candidates, {\em 2)} galaxies with a co-existence of BB and emission lines and {\em 3)} candidate Little-Red-Dot galaxies \citep[LRD, ][]{Labbe2023UNCOVER:-Candid, Matthee_2024_Little_Red_Dots, Greene_2024_UNCOVER_Spectroscopy} where a red continuum makes it difficult to established the presence of a Balmer break. Finally, another category includes spectra of insufficient S/N, precluding an evaluation of the BB.

We consider that all objects from Fig.~\ref{fig_spectrum_CLEAR_BB_objects} are in category 1. The object at $z=7.26$ shown in the top left panel is object JADES 8079 from V24, which strongly resembles the mini-quenched galaxy of L24. The origin of the flux excess at 4150 nm could be due to the \oiii\ lines, but the photometry is essentially constant longward of the BB.
Object $1181\_38684$ at $z=7.48$ (top right) shows again a clear BB, both in photometry and spectroscopy.
It exhibits a somewhat stronger break and a redder UV slope than the source of L24, possibly indicating an older age of its stellar population. 
For $1181\_74111$ at $z=8.53$, spectroscopy and photometry indicate a clear BB, although our photometric measurement using F277W and F410M shows a weak BB. The exact reason for the flux excess in F444W is not clear; it could be due to the presence of emission lines but the spectrum does not clearly support this hypothesis. 
Finally, the bottom left panel illustrates our highest-$z$ BB candidate ($1345\_80026$, $z=9.56$), with a well-established break from photometry, but an insufficient quality spectrum to firmly establish the origin of the break.
Overall, we count 5 objects with BB$>1.2$ where both spectroscopy and photometry show an essentially flat SED (in $F\nu$ units) longward of the break (including two objects from L24 and V24).

To further ensure the reliability of these BB candidates, we investigate the strengths of rest-optical lines and their possible contribution to the emission redward of the BB. We assume Gaussian profiles to fit simultaneously the H$\beta$ and [O~{\sc iii}] $\lambda\lambda$4960,5008 emission and a constant for the continuum (assumed here between $\lambda_{0} = 4660-5200$\,\AA). H$\beta$ is not detected in any of our strong BB candidates, with 2$\sigma$ upper limits between $EW_{0} (\rm H\beta) \leq 21.6$\,\AA{ }(for $1210\_5115$) and $\leq 137.8$\,\AA{ }(for $1345\_80026$). These are substantially lower than those observed in the averaged stacked spectra from \cite{Roberts-Borsani2024Between-the-Ext} ($\simeq 140-160$\,\AA{ }for their $z\simeq 7-9$ stacks). Therefore, the BB measurements of these objects should not be affected by emission lines.

Objects from categories 2 and 3 are shown in Fig.~\ref{fig_spectrum_objects}. The spectrum of $1345\_80372$ at $z=7.48$ shows \hb\ and \Oiii\ emission, which indicates that the F356W filter and hence our photometric measure of BB could be boosted by H$\gamma$ and other emission lines. On the other hand, a spectral break is quite clearly present in the spectrum, suggesting that this is probably an object showing both a Balmer break plus emission lines, similar to the object recently reported by \cite{Witten_2024_Rising_from_the_ashes_A2744-YD4}.
Finally, $2565\_15420$ at $z=7.14$ illustrates objects whose photometry and spectra resemble those of LRDs, with a blue/flat continuum shortward and a red continuum longward of the BB (and indications for emission lines in some cases), as shown e.g.~by \cite{Kocevski2024The-Rise-of-Fai}. In these cases, it is difficult to measure the true BB from photometry, but some studies have already reported the presence of a Balmer break in LRDs \citep[e.g.][]{Wang_2024_RUBIES_LRD_BB_z3, Wang_2024_RUBIES_Evolved_Stellar_Populations}.
From our inspection of the data, we consider that 5-7 galaxies show simultaneous signs of a Balmer break and emission lines, which most likely indicate the presence of renewed star-formation after some inactive period, aking to sequences of multiple bursts \citep{Witten_2024_Rising_from_the_ashes_A2744-YD4}. The spectra of these sources and the properties of the whole spectroscopic sample (main sample and the additional sample) are shown and tabulated in the Appendix (Fig.~\ref{fig_spectrum_additional_possible_BB_objects} and Table \ref{tab_properties_bb_cand}).

\begin{figure*}[htb]
    \centering
    \includegraphics[width=18cm]{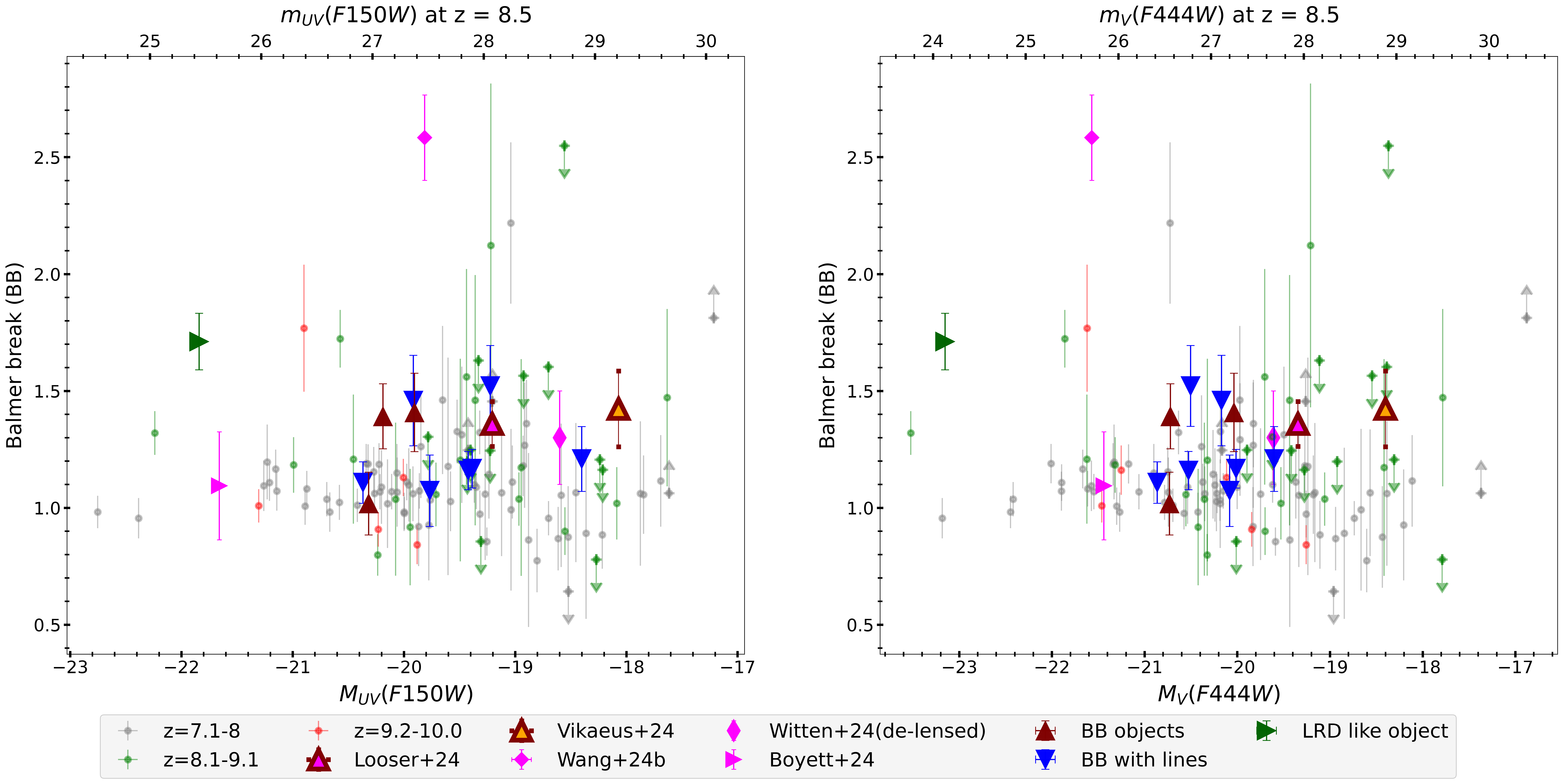}
    \caption{BB as a function of $M_{UV}$ (\textbf{left}) and $M_{V}$ (\textbf{right}) (in AB magnitude) for all spectroscopically confirmed galaxies with $z \sim 7.1-10$. The apparent magnitude of these sources at redshift 8.5 is also shown in the secondary x-axis.}
    \label{fig_bb_vs_mF444W}
\end{figure*}

\subsection{Balmer breaks are found from bright to faint sources}

To examine if sources showing strong Balmer breaks are preferentially found in bright/massive galaxies, we show in Fig.~\ref{fig_bb_vs_mF444W} our measurements of BB as a function of the rest-UV and V-band absolute magnitudes (probed by F150W and F444W, respectively). The 10-11 sources with BB, i.e.~either showing BB alone or BB with lines, are highlighted in this figure. No specific trend is seen for BB with either $M_{UV}$ or $M_{V}$, and objects with strong and/or secure BBs are found over a range of magnitudes, approximately from $M_{V}$ ($m_{F444W}) \sim$ -21 to -18 (24 - 29), i.e.~over two dex in rest-optical fluxes. This is in agreement with earlier studies which examined the behavior of the BB with rest-UV magnitude/luminosity \citep{Vikaeus_2024MNRAS_To_be_or_not_to_be,Roberts-Borsani2024Between-the-Ext, Wilkins_2024_FLARES_BB}.

The presence of significant BBs over a broad range of $M_{V}$ and $m_{F444W}$ indicates that BB galaxies are found for a variety of UV magnitudes, as also noticed earlier \citep[e.g.][]{Vikaeus_2024MNRAS_To_be_or_not_to_be}, and presumably also over a range of stellar masses. However, detailed SED fits are necessary to translate the observed properties into physical ones, e.g.~stellar masses, which will be presented in a detailed follow-up study (Kuruvanthodi et al, in prep). In any case, these observations suggest that galaxies with strong BB can exist over a larger range in masses, from a few times $10^8$ (e.g. L24) \msun\ to $> 10^9$ \msun, and even higher, e.g.~for LRD-type objects \citep[see e.g.][]{Wang_2024_RUBIES_Evolved_Stellar_Populations}.

\subsection{Balmer break galaxies at $z \protect\sim 7-9$ are more frequent than previously thought}
Out of the 101 sources for which secure spectroscopic redshift is available, we found spectra for 62 objects in DJA. Out of these 62, 5 ($\sim 8\%$) sources show a BB with a flat red continuum like in a mini-quenched galaxy, 5-7 ($\sim 10\%$) sources show a Balmer break with emission lines and one ($\sim 2\%$) is an LRD-like candidate with a possible break. That means that $\sim 16-20\%$ of the sources in our sample show a BB or the presence of an evolved stellar population.

Alternatively, an inspection of the spectra of all objects with photometric BB$>1.2$ (16 sources) yields the following: a BB is clearly detected and thus confirmed in 6 of them ($\sim 37$ \%), 6 others show emission lines and no clear break (37 \%), and the remaining spectra are of insufficient quality. This translates to a failure rate of 37\%  (maximum 63\%) to find true BB candidates from our photometric method.

If we consider the full photometric sample (with secure spectroscopic redshifts), 27 sources show BB above 1.2 out of 101 ($\sim 27$\%), comparable to or slightly higher than the inference from the spectra.
If one assumes the above-mentioned failure rate (and range thereof) for the whole sample, $\sim 10-17$ \% will have a true BB, which is larger than the previously confirmed BB objects but comparable to the fraction of ``smouldering'' (i.e.~in with suppressed star-formation) galaxies around at $z \sim 7-8$ determined by  \cite{Trussler_2024_Like_a_candle}.  
In any case, with a total sample of 10-11 spectroscopically confirmed BB objects at $z>7$, our work essentially triples the number of such objects.

Previously, JWST spectroscopic observations revealed 4 objects with significant Balmer breaks at $z > 7$ \citep{Looser2024A-recently-quen,Vikaeus_2024MNRAS_To_be_or_not_to_be, Wang_2024_RUBIES_Evolved_Stellar_Populations, Witten_2024_Rising_from_the_ashes_A2744-YD4} and a galaxy with a BB $\sim$ 1 \citep{Boyett_2024NatAs_A_massive_interacting}, as also shown in Figs.~\ref{fig_bb_vs_z_lit} and \ref{fig_bb_vs_mF444W} \footnote{One additional galaxy with spectroscopic redshift $z=7.10$ and BB$\sim 1.4$ is reported in the work of \cite{Trussler_2024_Like_a_candle} mentioned above.}. These studies reported individual objects selected by different criteria, and which do not result from a systematic search for strong BB objects. Their BB measurements from spectroscopy agree well (within 1-2 $\sigma$) with ours from photometry, as seen from  Fig.~\ref{fig_bb_vs_z_lit}, and the spectroscopic measurements are not affected by possible emission lines. Our sample not only adds new spectroscopically-confirmed BB objects, but it also extends somewhat the magnitude range over which such objects are found, as shown in \ref{fig_bb_vs_mF444W}.

\subsection{Origin of the observed Balmer breaks}
None of the objects reported here shows a Balmer break exceeding any fundamental limit. In particular, the observed values of BB are within the range of predicted for normal stellar populations and can be reached with ages less than the Hubble time, as shown earlier by \cite{Wang_2024_RUBIES_Evolved_Stellar_Populations} for the strongest BB object studied here. The BB strengths measured in this study are in agreement with the theoretical predictions of \citet{Binggeli_2019_Balmer_breaksinsimulated} and \cite{Wilkins_2024_FLARES_BB}, where some simulations predict BB strength up to 1.5 - 2.0 in redshift $\sim 10$ with extinction. 
Smaller values are predicted by \citet{Steinhardt_2024The_Highest_redshift_Balmer_Breaks} who use, however, D4000, which shows a small dynamical range due to chosen wavelength windows. In short, our strong BB candidates do not break standard cosmology, but they might indicate the need for new star formation models or obscured star formation \citep[see Fig. 5 in][]{Steinhardt_2024The_Highest_redshift_Balmer_Breaks}.

Most likely, the sample presented here represents the bursty nature of galaxies of $z \ga 7$ galaxies. Indeed, many theoretical and observational studies have already proposed that star formation in the early Universe might be bursty and/or stochastic 
\citep[e.g.][]{Caputi_2017_Star_Formation_in_Galaxiesat_z_4_5, Faisst_2019_The_Recent_Burstiness_of,Rinaldi_2022_The_Galaxy_Starburst,Endsley_2024_The_star_forming_and,Looser2023JADES:-Differin,Sun_2023_Seen_and_unseen_bursty_star_formation,Looser2024A-recently-quen,Witten_2024_Rising_from_the_ashes_A2744-YD4}.
These starburst phases are accompanied by temporary or permanent phases of quiescence, presumably due to various feedback mechanisms arising after the burst phase, which then allows the older stellar population (and the BB) to become visible in the integrated spectra. 
The co-existence of BBs with emission lines probably indicates the restart of star formation after a temporary quiescence \citep{Witten_2024_Rising_from_the_ashes_A2744-YD4}. 
In summary, the observed range of BB strengths and the existence of BB galaxies with emission lines indicate the galaxies at $z \ga 7$ are transiting between on and off phases of star formation \citep[cf.][]{Witten_2024_Rising_from_the_ashes_A2744-YD4, Looser2024A-recently-quen, Trussler_2024_Like_a_candle}. The presence of these galaxies provides constraints on the star-formation histories in the early Universe and their detailed modeling will help us to constrain the onset of galaxy formation.

A more detailed study of the evolution of the BB in large galaxy samples and over a wider redshift range will be presented in a follow-up study (Kuruvanthodi et al., in prep.).

\section{Conclusions}
\label{s_conclude}
To progress towards establishing the occurrence and properties of galaxies showing Balmer breaks (BBs) at $z>7$, we have selected all galaxies with secure spectroscopic redshifts from the main extragalactic surveys (CEERS, JADES, PRIMER, FRESCO) observed with JWST for which measure the strength of the Balmer break from photometry. We have validated our method by comparison with spectroscopic measurements of the BB from the stacked spectra of \cite{Roberts-Borsani2024Between-the-Ext}, and we have inspected the individual spectra to validate our measurements and assess the nature of the galaxies.

Our work has revealed three new galaxies from $z \sim 7.1-9.6$ with strong BBs and no signs of emission lines, which have most likely stopped star formation recently (so-called mini-quenched objects).  We have also found 5-7 galaxies showing the co-existence of strong BBs and emission lines (\hb, \oiii), which we interpret as signs of renewed star formation after a previously interrupted (quenched) phase. This approximately triples the number of spectroscopically-confirmed galaxies at $z>7$ showing clear signs of quenched or interrupted star formation, with respect to previous studies \citep{Looser2024A-recently-quen,Vikaeus_2024MNRAS_To_be_or_not_to_be, Wang_2024_RUBIES_Evolved_Stellar_Populations, Witten_2024_Rising_from_the_ashes_A2744-YD4}.

Overall, from our sample of 118 galaxies, 101 yielded BB measurements, with approximately 48\% showing a net BB, and $\sim 27$\% a ``strong'' BB$>1.2$. From a detailed inspection of the available spectra (for a subsample of 62 objects), we estimate that $\sim 20$\% show a clear BB, out of which approximately half show no signs of ongoing star formation, and the rest show BB plus emission lines. The remaining objects with apparent strong BBs resemble LRDs or cannot well be classified.
Our photometric BB measurements show that objects with strong BBs can be found over a wide range of magnitudes and stellar masses. From our analysis, we conclude that the fraction of mini-quenched or reactivated (bursty) galaxies at $z>7$ is of the order of $\sim 10 - 20$\%, at least in the sample studied here. 

Our sample, showing a range of BB strength and line properties, highlights the bursty and stochastic nature of star formation in the early Universe and the existence of significant galaxies in a temporary phase of quiescence in the high-z Universe. Future work will be needed to understand the possible selection biases, establish how these trends evolved towards lower redshift, quantify the timescales involved in the observed star formation and intermittent phases, and work out more general implications.

\begin{acknowledgements}
We thank the anonymous referee for her/his insightful comments and suggestions. The data products presented herein were retrieved from the Dawn JWST Archive (DJA). DJA is an initiative of the Cosmic Dawn Center (DAWN), which is funded by the Danish National Research Foundation under grant DNRF140. We thank Gabe Brammer for making this tool available. We thank Prof. Corinne Charbonnel for her useful comments and suggestions. This work has received funding from the Swiss State Secretariat for Education, Research, and Innovation (SERI) under contract number MB22.00072, as well as from the Swiss National Science Foundation (SNSF) through project grant 200020\_207349

\end{acknowledgements}
\bibliographystyle{aa}
\bibliography{bb_jwst_v2}

\begin{appendix}
\section{Spectra of other Balmer break candidates}
\label{spectra_additional_bb_lines}
Spectra of other BB candidates with emission lines are shown here. 
\begin{figure*}[htbp]
    \centering
    \includegraphics[width=8.7cm]{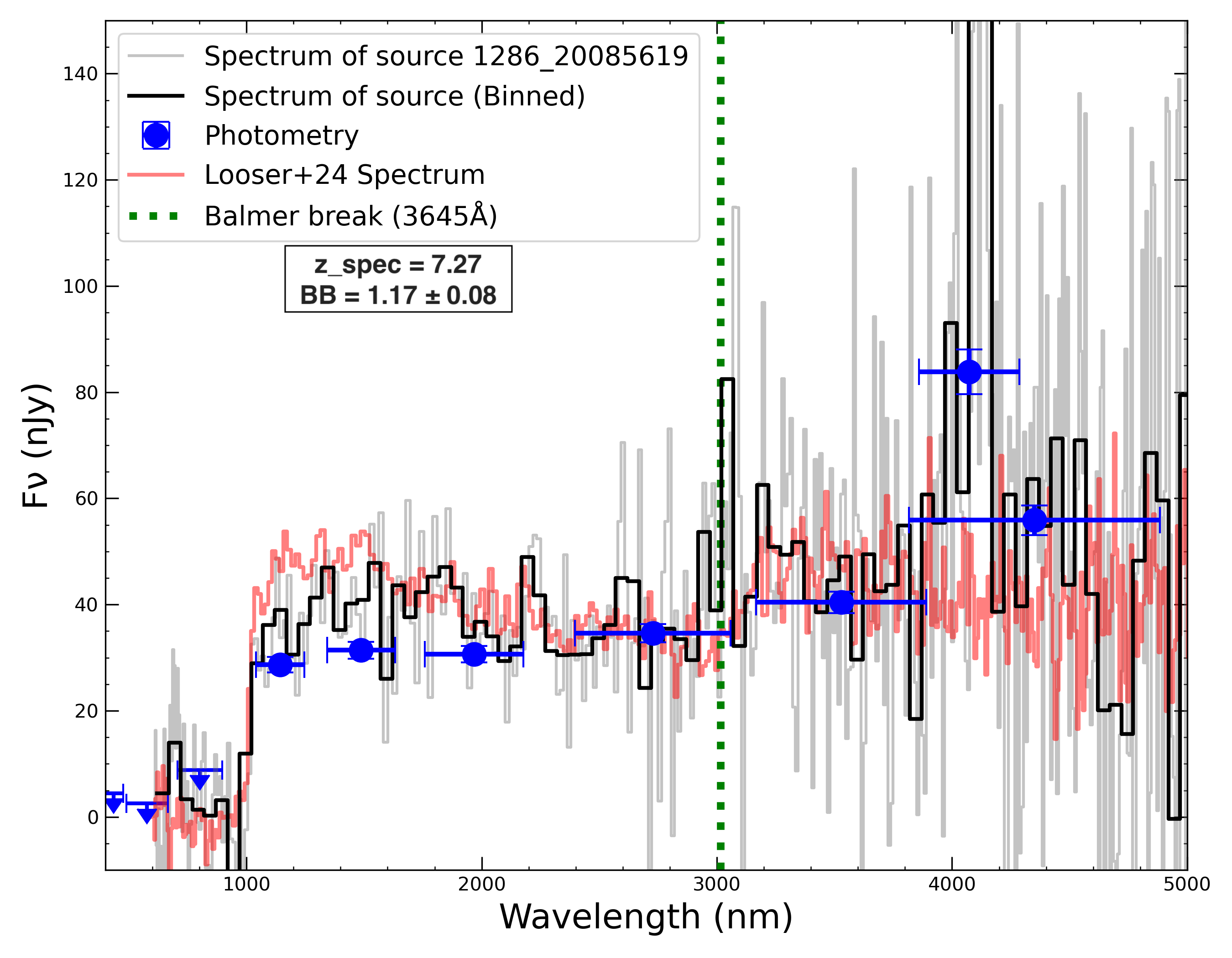}
    \includegraphics[width=8.7cm]{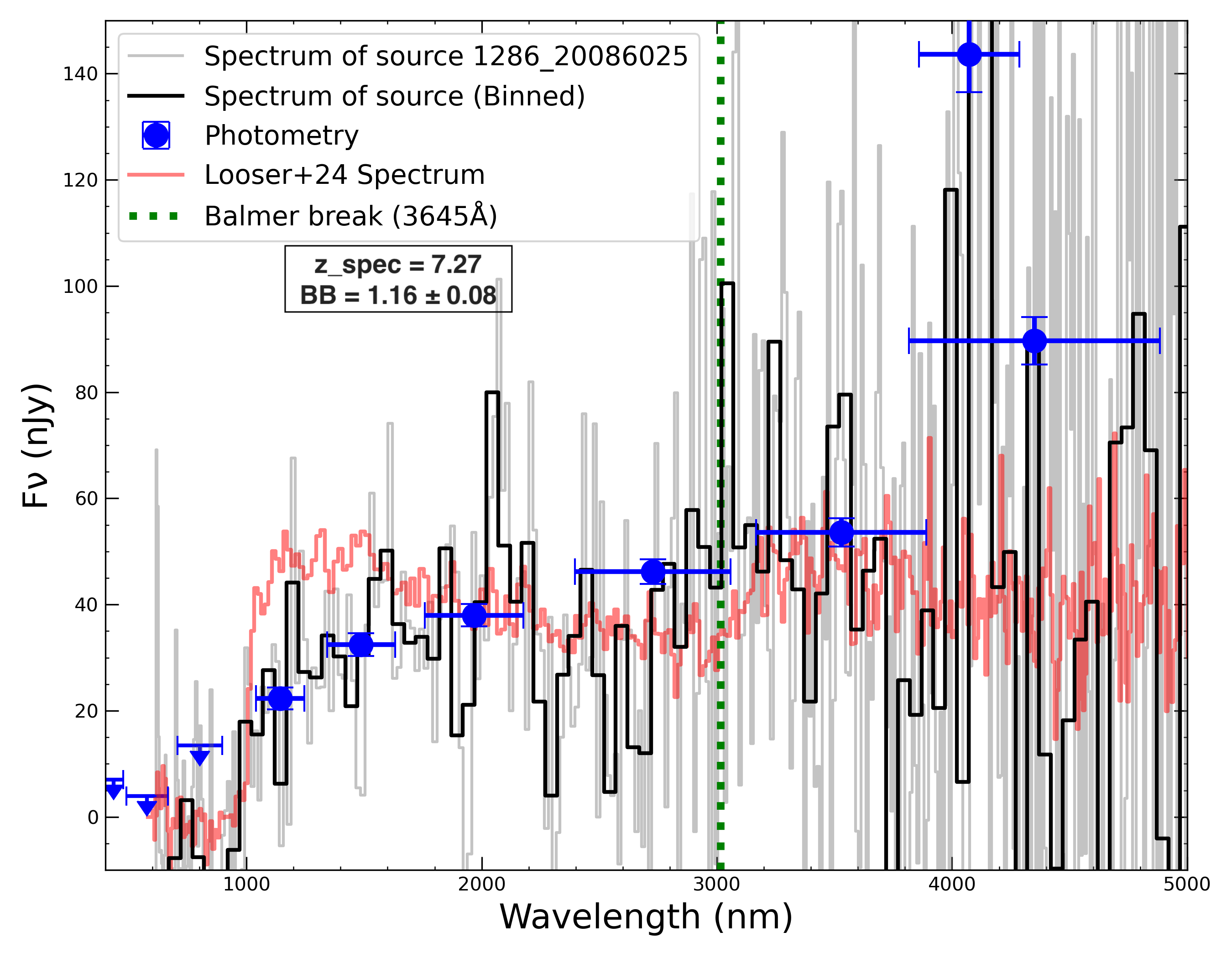}
    \includegraphics[width=8.7cm]{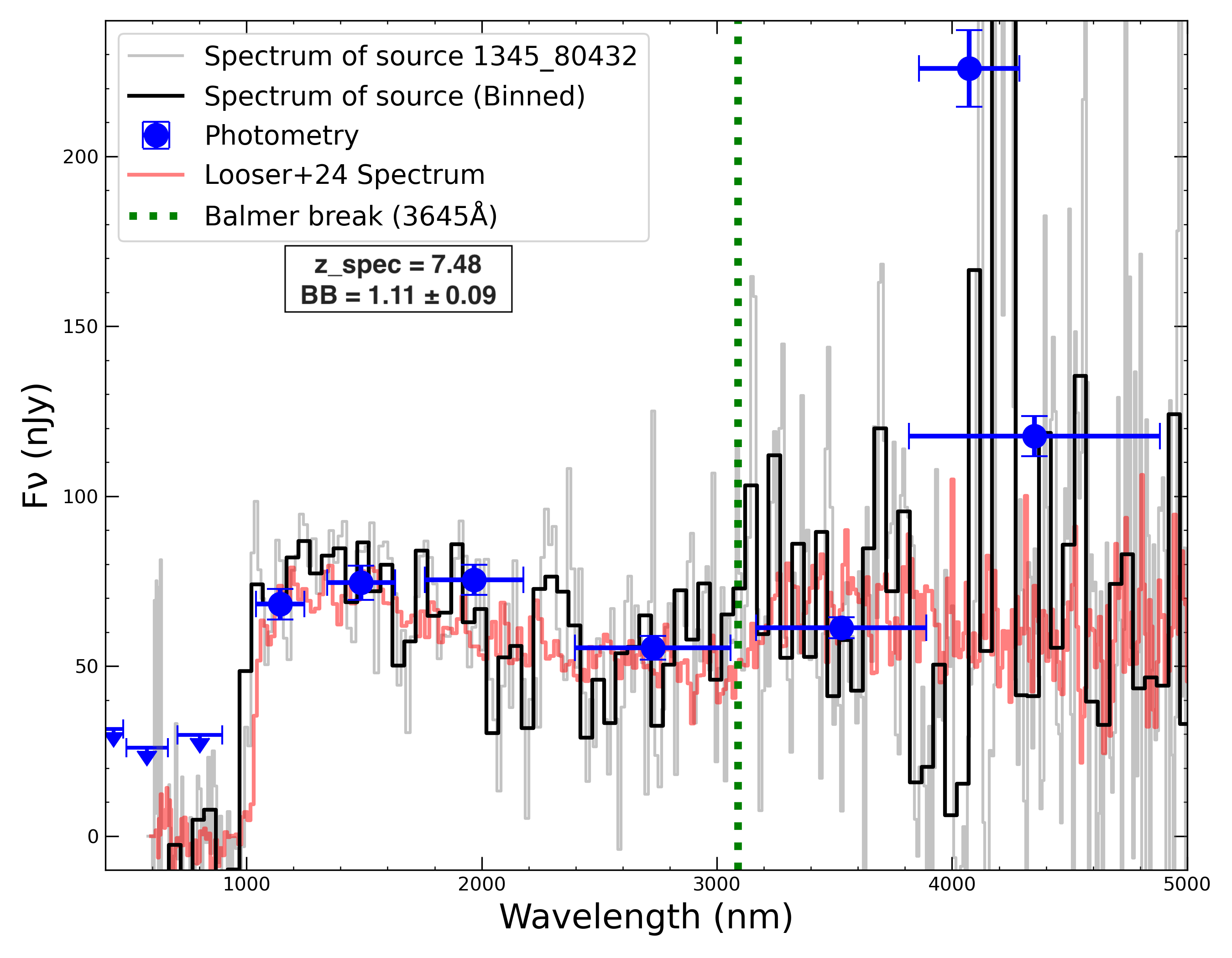}
    \includegraphics[width=8.7cm]{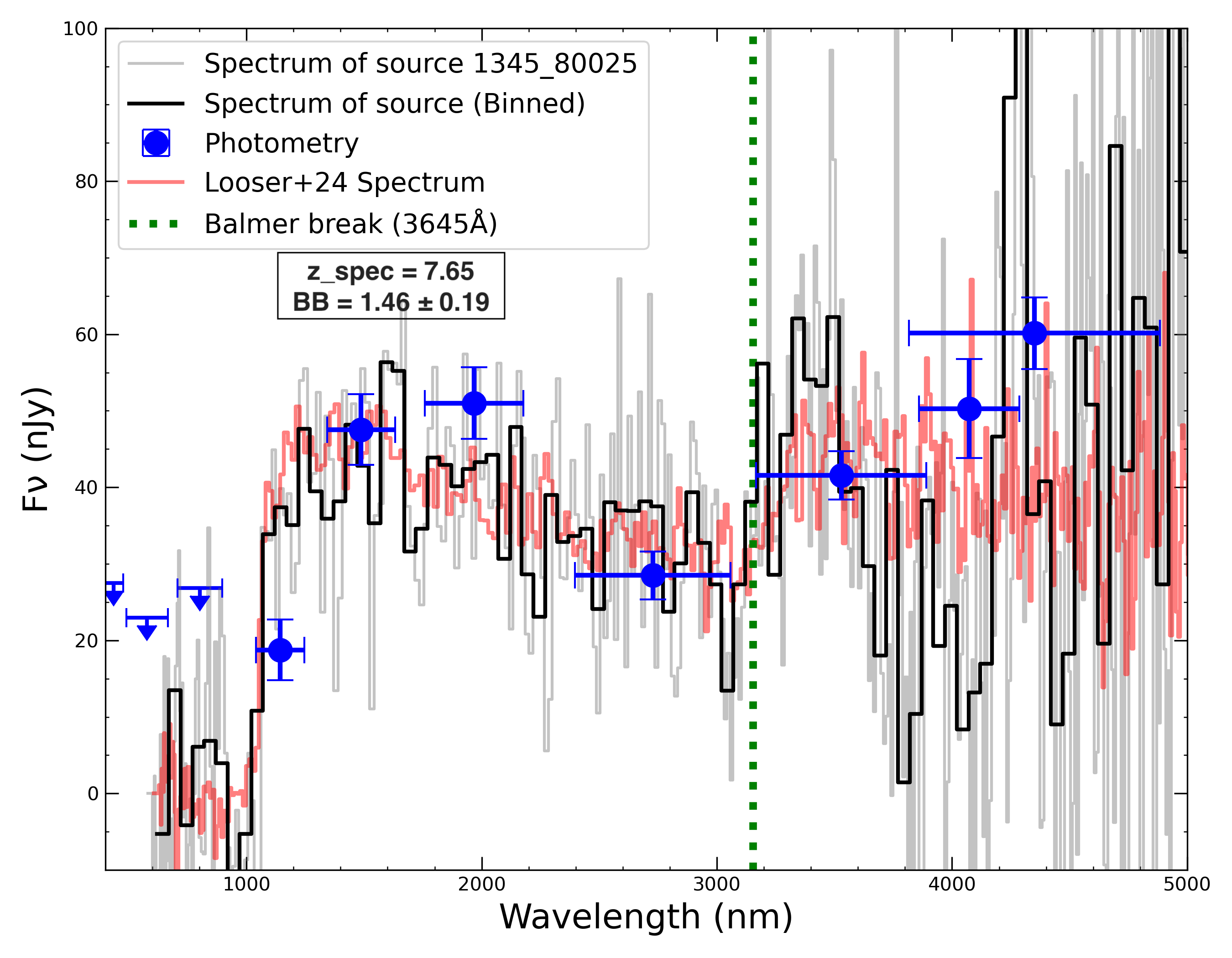}
    \includegraphics[width=8.7cm]{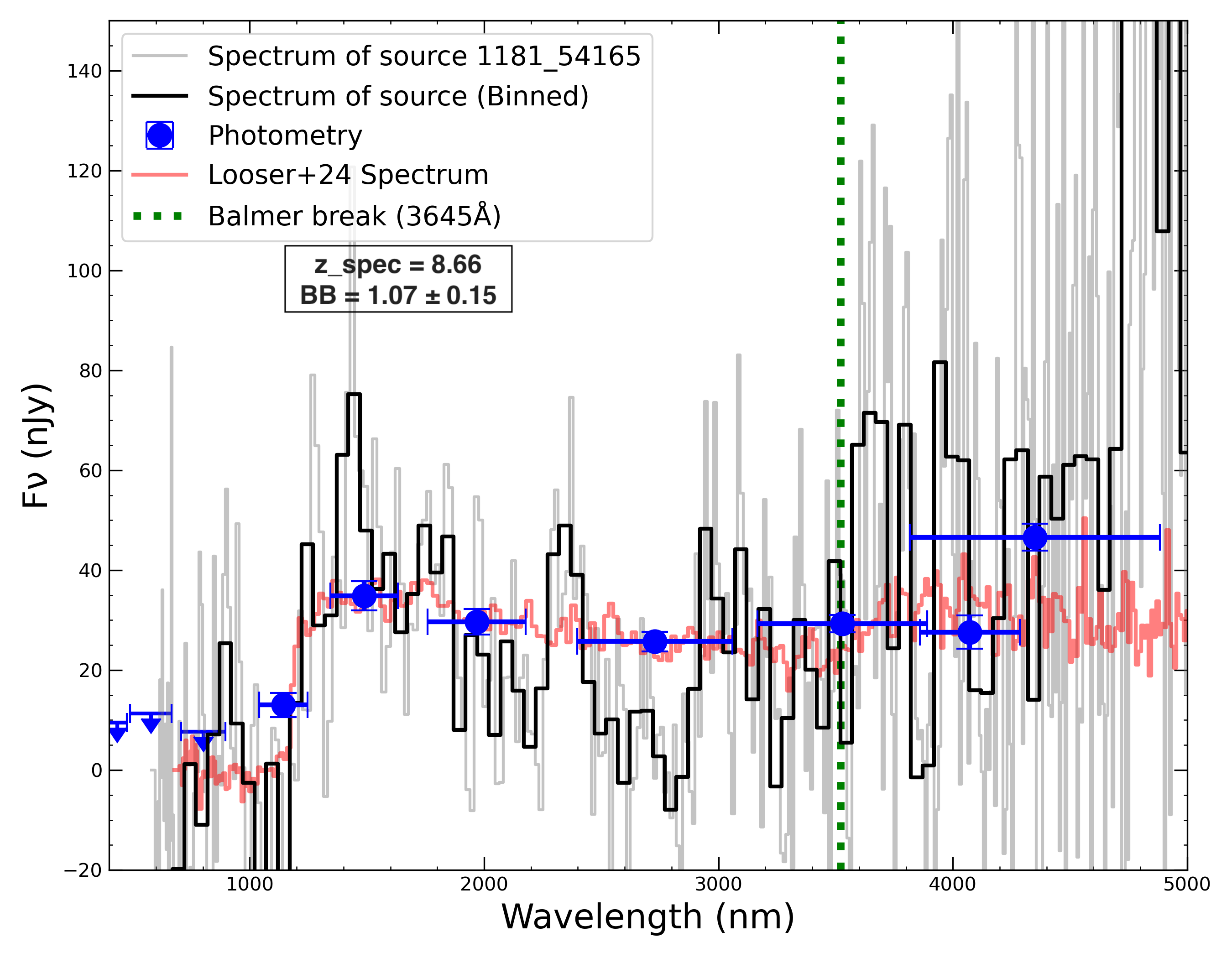}
    \includegraphics[width=8.7cm]{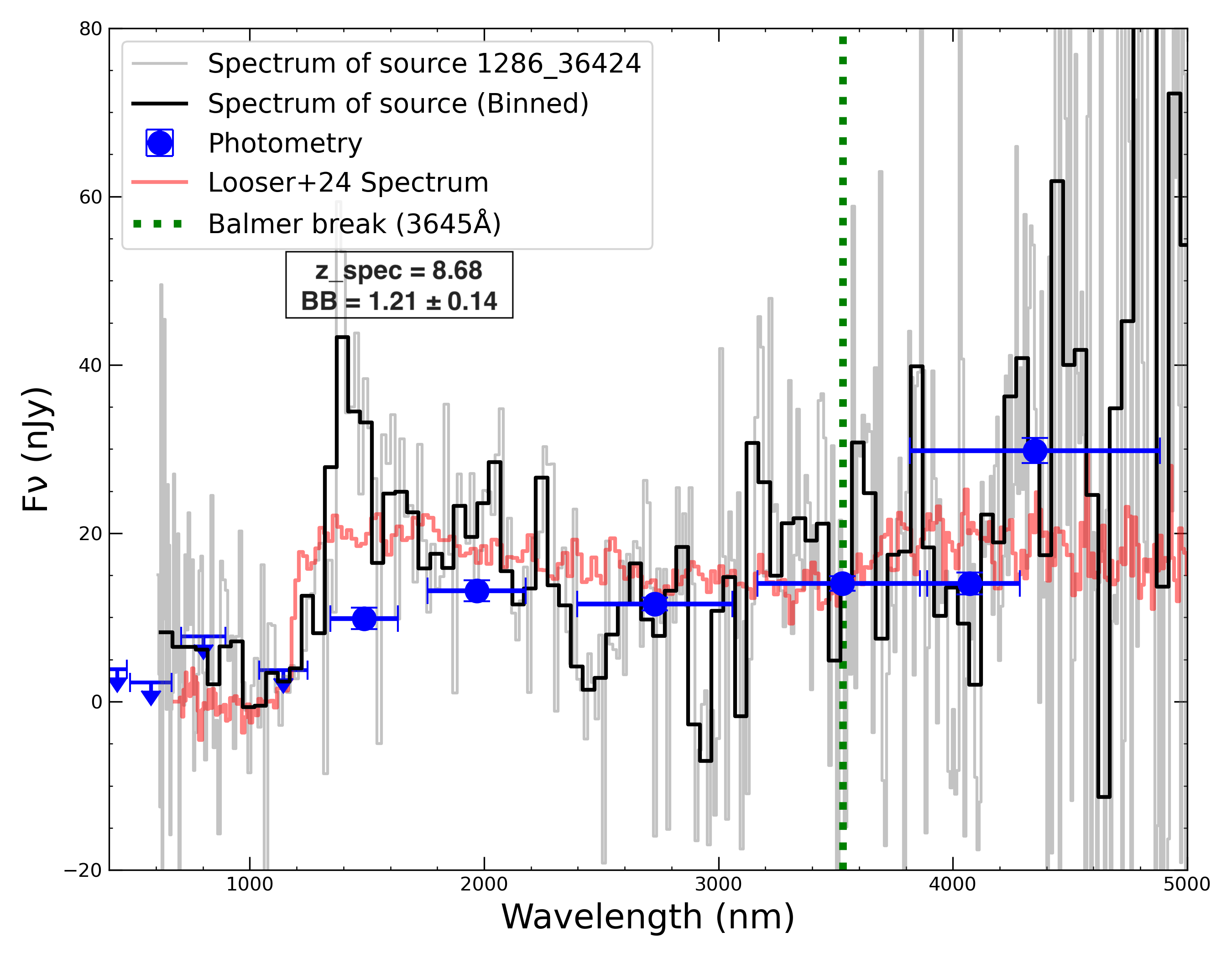}
    \caption{NIRSpec PRISM spectra of additional BB + emission line candidates at $z \sim 7-9.5$, showing objects with clear or plausible Balmer breaks. Binned and unbinned spectra are shown for illustration. For comparison, we also over-plot the spectrum of the mini-quenched $z=7.3$ galaxy from \cite{Looser2024A-recently-quen}, redshifted and rescaled to approximately match the observed flux blueward of the BB.}
    \label{fig_spectrum_additional_possible_BB_objects}
\end{figure*}\\

\section{Coordinates and properties of the Balmer break sources/candidates}
\label{properties_of_BB_cands}
\begin{table*}[ht]
    \centering
    \caption{Properties of BB candidates. The top panel shows the 6 BB galaxies discussed in the main text and the bottom panel shows the additional BB candidates with emission lines.}
    \begin{tabular}{ccccccc}
    \hline
    \hline
    ID  & $z_{spec}$ &  RA  & Dec & BB & $m_{F444W}$ & $EW(H_{\beta})$ \\
        &            & (deg) & (deg) &    &    (mag)   &  (\AA) \\
    \hline
    $1210\_5115$ & 7.26 & 53.1528407 & -27.8019455 & $1.42\pm0.16$ & $28.64\pm0.08$ & <21.55 \\
    $1181\_38684$ & 7.48 & 189.1211092 & 62.2778222 & $1.41\pm0.17$ & $27.05\pm0.09$ & <59.36 \\
    $1181\_74111$ & 8.53 & 189.1807727 & 62.1805329 & $1.02\pm0.13$ & $26.56\pm0.05$ & <36.52 \\
    $1345\_80026$ & 9.56 & 214.8118531 & 52.7371100 & $1.39\pm0.14$ & $26.74\pm0.07$ & <137.80 \\
    $1345\_80372$ & 7.48 & 214.9278186 & 52.8500015 & $1.52\pm0.17$ & $26.59\pm0.05$ & $130.39\pm21.56$ \\
    $2565\_15420$ & 7.14 & 150.0990078 & 2.3436228 & $1.71\pm0.12$ & $23.86\pm0.05$ & $67.43\pm14.96$ \\
        &   &   &   &   &   &   \\
    \hline
        &   &   &   &   &   &   \\
    $1286\_20085619$ & 7.27 & 53.1910569 & -27.7973172 & $1.17\pm0.08$ & $27.03\pm0.05$ & $47.70\pm16.05$ \\
    $1286\_20086025$ & 7.27 & 53.1837525 & -27.7938918 & $1.16\pm0.08$ & $26.52\pm0.05$ & <148.60 \\
    $1345\_80432$ & 7.48 & 214.8120544 & 52.7467436 & $1.11\pm0.09$ & $26.22\pm0.05$ & $229.54\pm59.55$ \\
    $1345\_80025$ & 7.65 & 214.8060747 & 52.7508628 & $1.46\pm0.19$ & $26.95\pm0.08$ & $70.58\pm29.90$ \\
    $1181\_54165$ & 8.66 & 189.2718353 & 62.1951783 & $1.07\pm0.15$ & $27.23\pm0.06$ & <31.61\\
    $1286\_36424$ & 8.68 & 53.1760306 & -27.8173945 & $1.21\pm0.14$ & $27.71\pm0.05$ & <72.98\\
    \hline
    \end{tabular}
\label{tab_properties_bb_cand}
\end{table*}
\end{appendix}

\end{document}